%% file: main.tex
\documentclass[journal]{IEEEtran}

\usepackage{cite}
\usepackage{array}
\usepackage{color}
\usepackage{float}
\usepackage[cmex10]{amsmath}
\usepackage{nomencl}						%nomenclature
\usepackage[normalem]{ulem}			        %underline
\usepackage{diagbox}						%table head
\usepackage{slashbox}						%table head
\usepackage{colortbl}						%table color
\usepackage{multirow}						%table combine
\usepackage{tabularx}
\usepackage{placeins}
\usepackage{algorithm}
\usepackage{mathrsfs}
\usepackage{mathdots}
\usepackage{amssymb}
\usepackage{dblfloatfix}
\usepackage{amsthm}
\usepackage{arydshln}
\usepackage{color}
\usepackage[noend]{algpseudocode}
\usepackage{bm}
\usepackage{url}
\usepackage{threeparttable}
\usepackage[hidelinks]{hyperref}
\usepackage[T1]{fontenc}
\newcommand{\subparagraph}{}
\usepackage{fancyhdr} 
\usepackage{extarrows}
\usepackage{enumitem}
\usepackage{booktabs}

\usepackage{booktabs,siunitx,makecell}
\usepackage{pifont}
\usepackage{capt-of}

\makeatletter
\newcommand{\slightlysmaller}{\fontsize{9.5}{11}\selectfont}
\makeatother

\AtBeginDocument{
\everymath{\slightlysmaller}
\everydisplay{\slightlysmaller}
}

\ifCLASSINFOpdf
  \usepackage[pdftex]{graphicx}
	\graphicspath{{./figure/}}
  \DeclareGraphicsExtensions{.pdf,.jpeg,.png}
\else
  \usepackage[dvips]{graphicx}
  \graphicspath{{./figure/}}
  \DeclareGraphicsExtensions{.eps}
\fi

\ifCLASSOPTIONcompsoc
  \usepackage[caption=false,font=normalsize,labelfont=sf,textfont=sf]{subfig}
\else
  \usepackage[caption=false,font=footnotesize]{subfig}
\fi

\newcommand{\figref}[1]{\figurename~\ref{#1}}
\newcommand{\tabref}[1]{Table~\ref{#1}}

\ifCLASSOPTIONonecolumn

\else

\renewcommand{\arraystretch}{1.5}
\fi

\makeatletter
\IEEEtriggercmd{\reset@font\normalfont\fontsize{7pt}{8pt}\selectfont}
\makeatother
\IEEEtriggeratref{1}

\begin{document}
%\bstctlcite{IEEEtran.bst}
\setcounter{page}{1}
\vspace{-4\baselineskip}
\title{Unified Energy Function Tailored to Inverter-Based Resources with PI Controllers for Transient Stability Analysis}
\author{Yifan~Zhang, \IEEEmembership{Student Member, IEEE}, Hsiao-Dong Chiang, \IEEEmembership{Fellow, IEEE}, \\ Yitong~Li, \IEEEmembership{Member, IEEE}, Yang~Wu, \IEEEmembership{Member, IEEE}
%\thanks{This work was supported by .)
%}
}
\IEEEaftertitletext{\vspace{-2.5\baselineskip}}
\ifCLASSOPTIONpeerreview
	\maketitle %\IEEEpeerreviewmaketitle
\else
	\maketitle
\fi
\thispagestyle{fancy}
\lhead{IEEE TRANSACTIONS ON POWER ELECTRONICS}
\rhead{\thepage}
\cfoot{}
\renewcommand{\headrulewidth}{0pt}
\pagestyle{fancy}

\begin{abstract}
The increasing penetration of inverter-based resources (IBRs) has fundamentally altered the transient stability characteristics of modern power systems. IBRs typically rely on proportional--integral (PI) controllers for synchronization and regulation, resulting in nonlinear swing equations that differ significantly from those of synchronous generators (SGs) and exhibit state-dependent damping. Consequently, although the classical energy function is often adopted in IBR analysis by analogy with SGs, it cannot be directly applied to IBRs with PI controller. A new \emph{energy function} explicitly tailored to \emph{PI controller} is proposed in this letter. It admits a unified form and can be applied to a class of nonlinear systems with PI controllers. 
Two representative cases are considered, including a grid-following (GFL) inverter and a DC-voltage-controlled grid-forming (GFM) inverter, demonstrating less conservative and more effective estimation of the region of attraction (ROA). All findings are verified through hardware-in-the-loop (HIL) experiments.
\end{abstract}
\bstctlcite{BSTcontrol}


\input{paper}
\input{appendix}

\ifCLASSOPTIONcaptionsoff
  \newpage
\fi
%\vspace{-0.1 in}
\bstctlcite{BSTcontrol}
\bibliographystyle{IEEEtran}
\bibliography{References}

\end{document}

%% file: paper.tex
\vspace{-12pt}
\section{Introduction}
\vspace{-3pt}
Electric power systems are undergoing a fundamental transition from being dominated by synchronous generators (SGs) to inverter-based resources (IBRs) as renewable generation continues to expand. The transient synchronization stability of IBRs—i.e., the ability to maintain or regain synchronization with the bulk grid after large disturbances—has therefore attracted increasing attention in recent years \cite{Xiongfei_overview, Gu_proceeding}. Among available assessment methods, the direct method based on Lyapunov/energy functions has been widely adopted due to its mathematical rigor, clear physical interpretation, and ability to provide conservative stability criteria \cite{chiang2011direct}. By estimating the region of attraction (ROA), this direct method enables transient stability assessment without extensive time-domain simulations or repeated critical clearing time (CCT) searches.

Recent studies have attempted to extend energy-function-based direct methods to IBR systems, including grid-forming (GFM) inverters \cite{shuai2018transient, fu2020large} and grid-following (GFL) inverters \cite{fu2020large, hu2019large, mansour2021nonlinear, zhang2021large}. Unlike SGs, IBRs achieve synchronization through software-based control loops, whose phase-angle outputs directly determine post-fault synchronism.  For GFM inverters, synchronization dynamics can resemble those of SGs when a low-pass filter (LPF) is introduced in the control loop \cite{d2013equivalence}. By contrast, GFL inverters synchronize with the point of common coupling (PCC) voltage through phase-locked loops (PLLs) implemented with proportional–integral (PI) controllers \cite{Xiongfei_overview, li2022revisiting}. However, energy-function construction is nontrivial due to control-based synchronization, particularly with PI controllers.

By deriving an equivalent swing equation for GFL inverters, \cite{hu2019large} constructed a classical energy function analogous to that of SG systems. However, the state-dependent damping induced by PLL nonlinearities may become negative in certain regions, leading to non-conservative stability evaluation \cite{fu2020large, zhang2021large}. Although numerical Lyapunov functions can also be computed using sum-of-squares \cite{zhang2020synchronizing} or iterative methods \cite{10032630}, they offer limited physical insight and poor scalability. Consequently, for IBRs with PI-controlled synchronization, the conventional swing-equation-based energy function yields an indefinite dissipation term, preventing direct ROA estimation through a critical energy level set.

PI-controlled synchronization structures are also being adopted in emerging IBR designs. A representative example is the DC-voltage-controlled GFM inverter -- also referred to matching control \cite{8825565} or frequency-following voltage-forming inverter \cite{Ai2024VFM}. This converter enables GFM operation with power-tracking capability through DC-voltage regulation and has emerged as a promising IBR configuration \cite{Zhao2022VFM}. Existing transient stability studies of this converter mainly rely on phase-portrait analysis \cite{Zhao2022VFM, luo2023design}, which does not provide quantitative ROA characterization. Although \cite{9762041} compared LPF- and PI-controlled GFM structures by deriving swing-equation models, the resulting energy dissipation remains indefinite because of the PI controller.

This letter addresses the above gap by developing a unified \emph{PI-controller energy function}. By incorporating the integrator-state energy into the Lyapunov construction, the proposed formulation eliminates the dependence on state-dependent damping and yields a unified energy function with a globally non-positive derivative. Consequently, the unstable equilibrium point (UEP) energy can be directly used as a valid level set for ROA estimation, enabling less conservative transient stability assessment for PI-controlled IBRs.
\begin{figure}[b!]
\vspace{-18pt}
\centering
\subfloat[LPF controller.]{\includegraphics[width=0.238\textwidth]{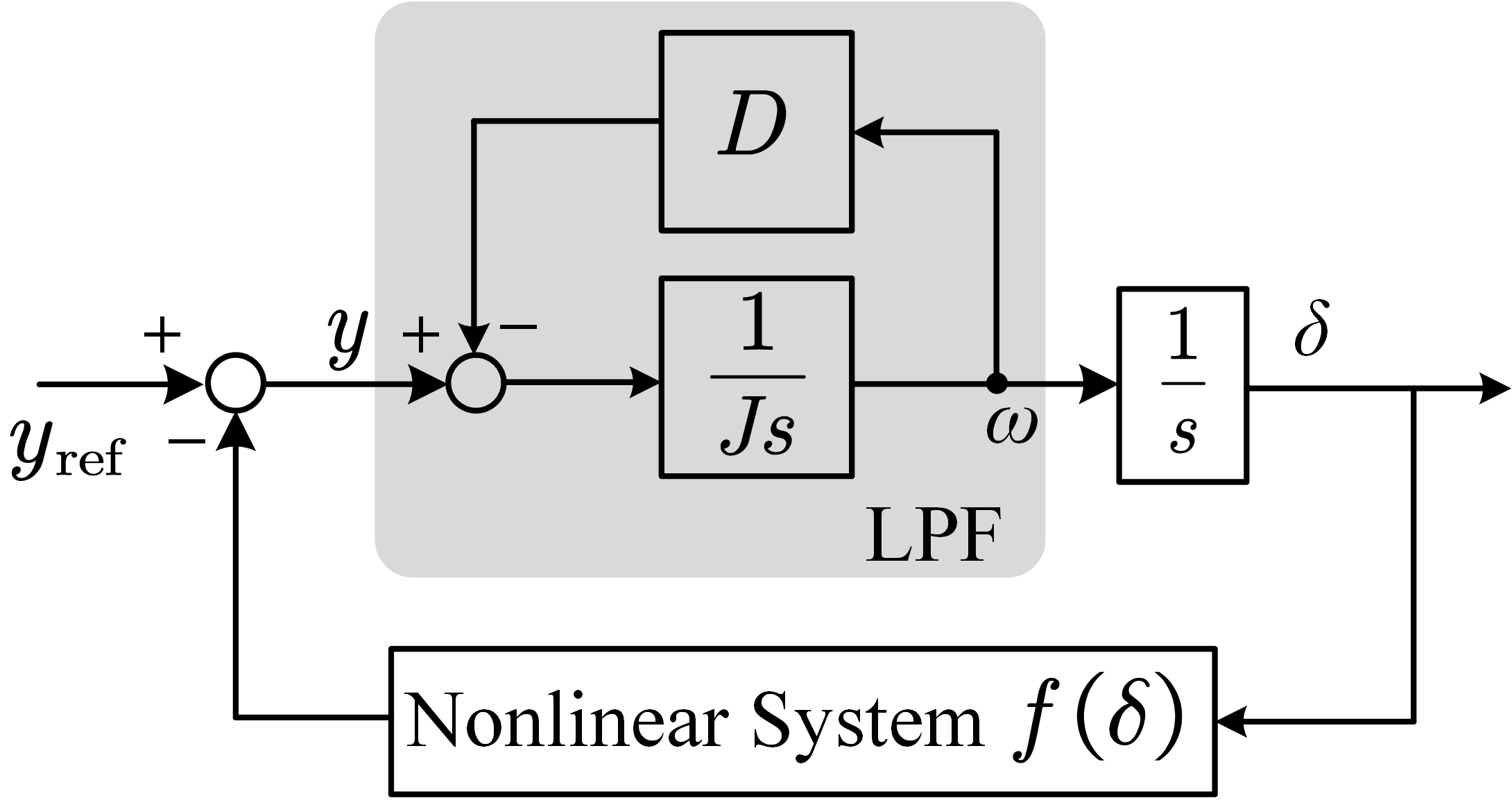}}\,
\subfloat[PI controller.]{\includegraphics[width=0.238\textwidth]{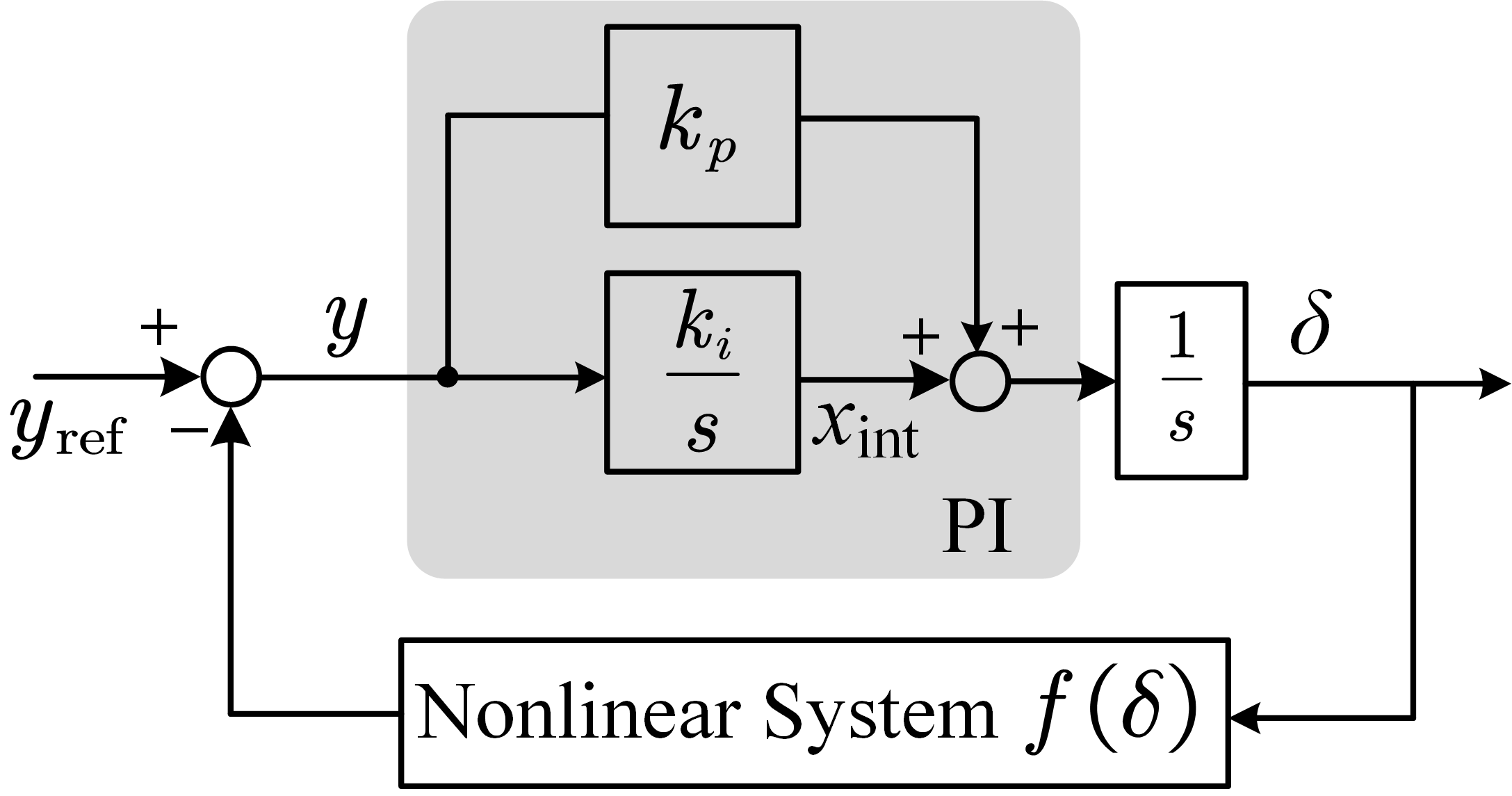}} 
\vspace{-3pt}
\caption{Control block diagram of the nonlinear system.}
\label{figcontroller}
\vspace{-3pt}
\end{figure}
\vspace{-3pt}
\section{Unified PI-Controller Energy Function}
\begin{figure*}[b!]
    \centering
    \subfloat[Grid-connected inverter system.]{\includegraphics[width=0.35\textwidth]{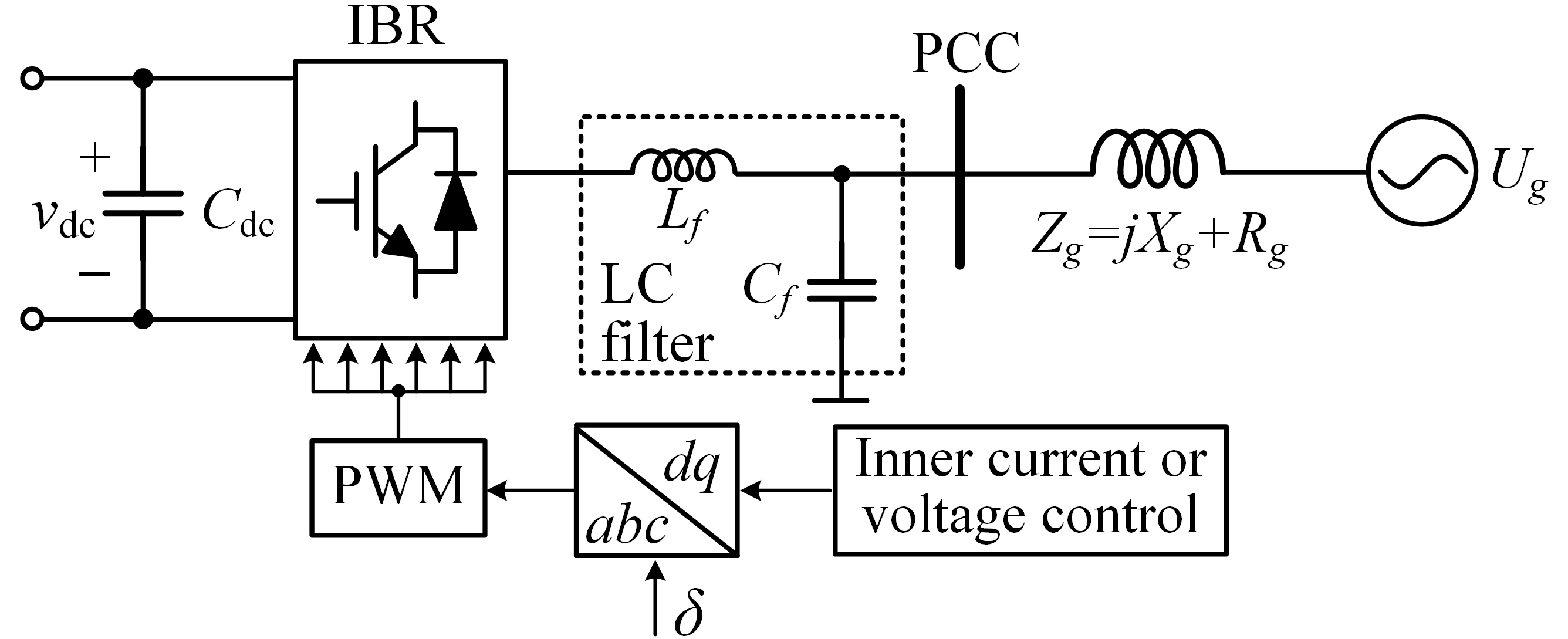}}
    \subfloat[PLL in GFL inverter.]{\includegraphics[width=0.3\textwidth]{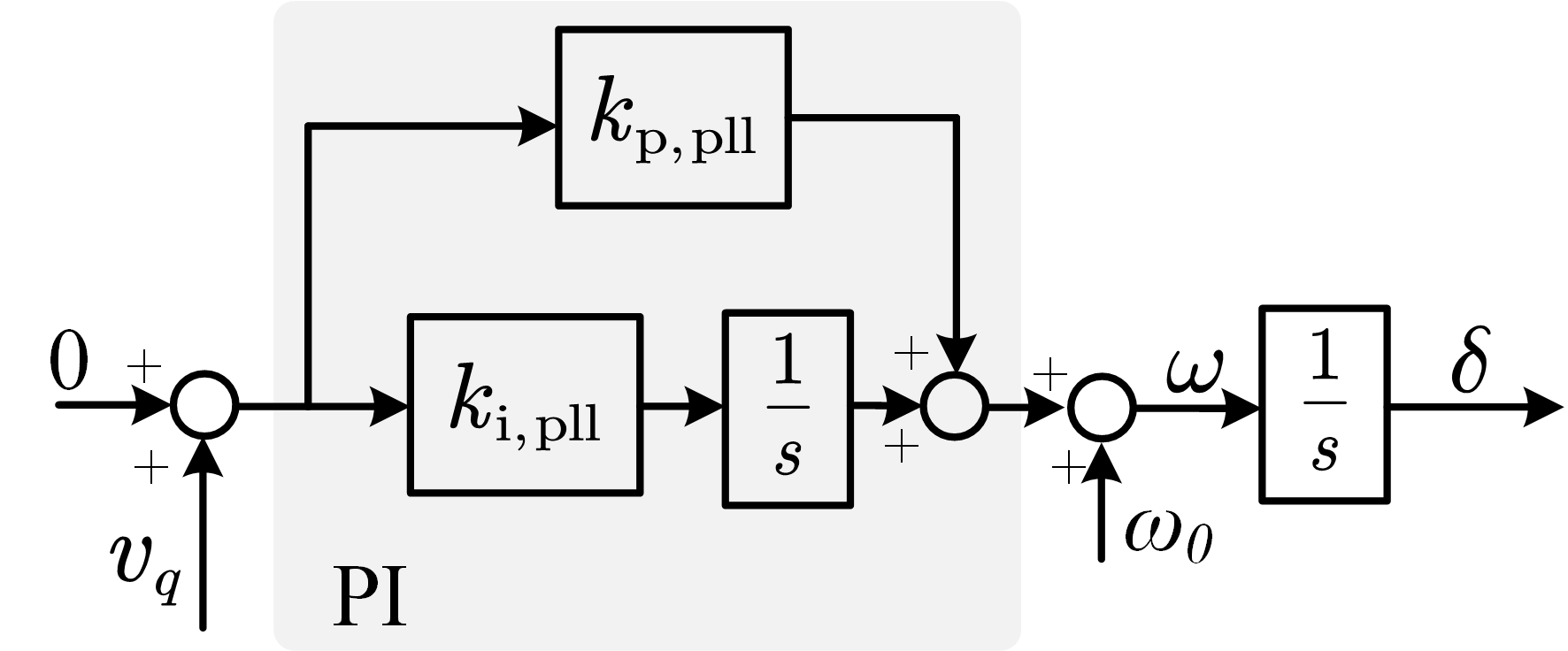}}
    \subfloat[DC-voltage-controlled GFM inverter.]{\includegraphics[width=0.34\textwidth]{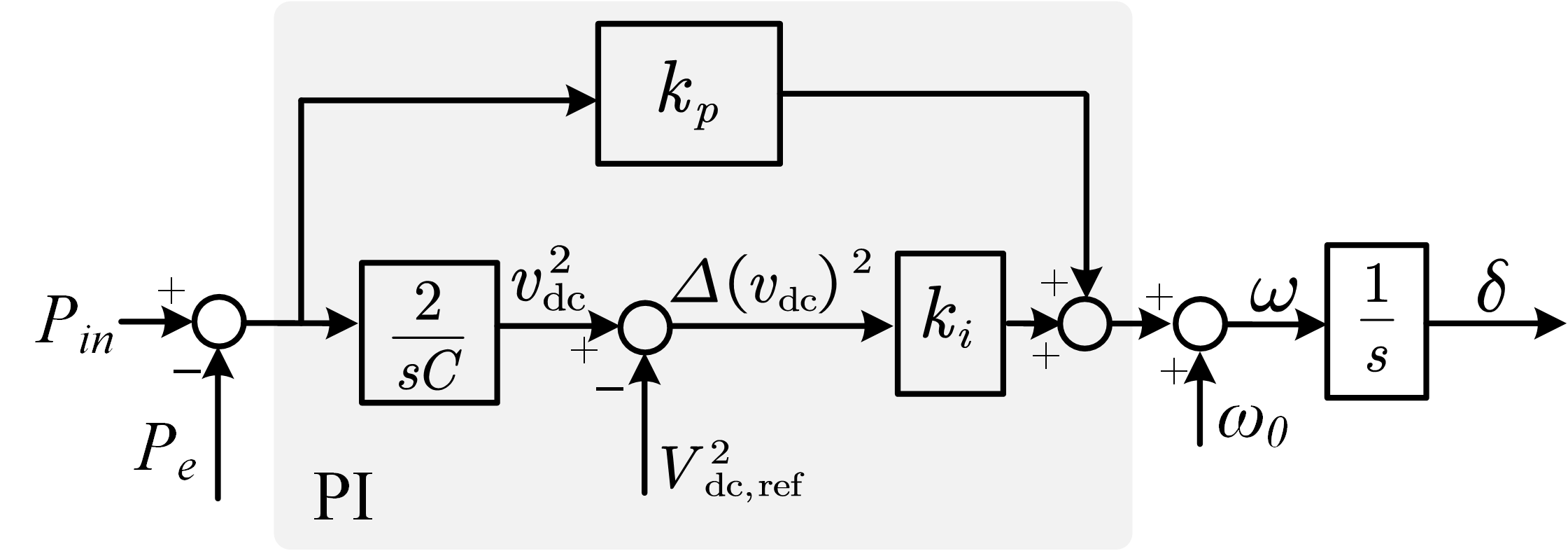}}
    \caption{Grid-connected inverter system and synchronization control loops.}
    \label{ST}	
\end{figure*}

In general, the nonlinear system with feedback controller can be abstracted to \figref{figcontroller}, which is also known as Lur'e system \cite{nonlinearbook_Khalil}. Two mainstream controllers are considered here: the low-pass filter (LPF) controller and the proportional–integral (PI) controller, shown in \figref{figcontroller}~(a) and (b), respectively. The nonlinear system is represented by $f(\delta)$.

For the LPF controller, the system can be readily expressed in a swing-equation form. The LPF time constant and gain act as equivalent inertia and damping, respectively. Specifically, the system in \figref{figcontroller}~(a) can be written
\begin{equation}
\vspace{-3pt}
    \begin{cases}
	\dot{\delta}=\omega\\
	J\dot{\omega}=\left[ y_{\mathrm{ref}}-f\left( \delta \right) \right] -D\omega\\
\end{cases}
\end{equation}
where $y_{\mathrm{ref}}-f(\delta)$ serves as the synchronizing torque. The LPF therefore naturally realizes damping and inertia, and the corresponding classical energy function is 
\begin{equation}
\vspace{-3pt}
W=\underset{\text{kinetic\, energy}}{\underbrace{\frac{1}{2}J\omega ^2}}\underset{\text{potential\,energy}}{\underbrace{-\int{\left[ y_{\mathrm{ref}}-f\left( \delta \right) \right] d\delta}}}
\label{EF_LPF}
\vspace{-3pt}
\end{equation}
with derivative 
\begin{equation}
\vspace{-3pt}
\dot{W}=-D\omega ^2
\end{equation}
\begin{table}[t!]
\centering
\vspace{-6pt}
\caption{Duality between Classical \& PI-Controller Energy Functions}
\vspace{-6pt}
\label{tab:duality}
\setlength{\tabcolsep}{2pt}
{\footnotesize
\everymath{}        
\everydisplay{}
\begin{tabular}{c|c|c}
\hline\hline
 & Classical (LPF-based) & Proposed (PI-based) \\
\hline
System structure 
& \figref{figcontroller}~(a)
& \figref{figcontroller}~(b) \\
\hline
Quadratic term 
& $\tfrac{1}{2}J\omega^2$ 
& $\tfrac{1}{2k_i}x_{\text{int}}^2$ \\
\hline
Potential energy
& $-\int (y_{\mathrm{ref}}-f(\delta))\,d\delta$ 
& $-\int (y_{\mathrm{ref}}-f(\delta))\,d\delta$ \\
\hline
Dissipation 
& $-D\omega^2$ 
& $-k_p(y_{\mathrm{ref}}-f(\delta))^2$ \\
\hline
Equivalent circuit 
& \begin{minipage}[t]{0.17\textwidth}
\centering
\vspace{-6pt}
\includegraphics[width=\linewidth]{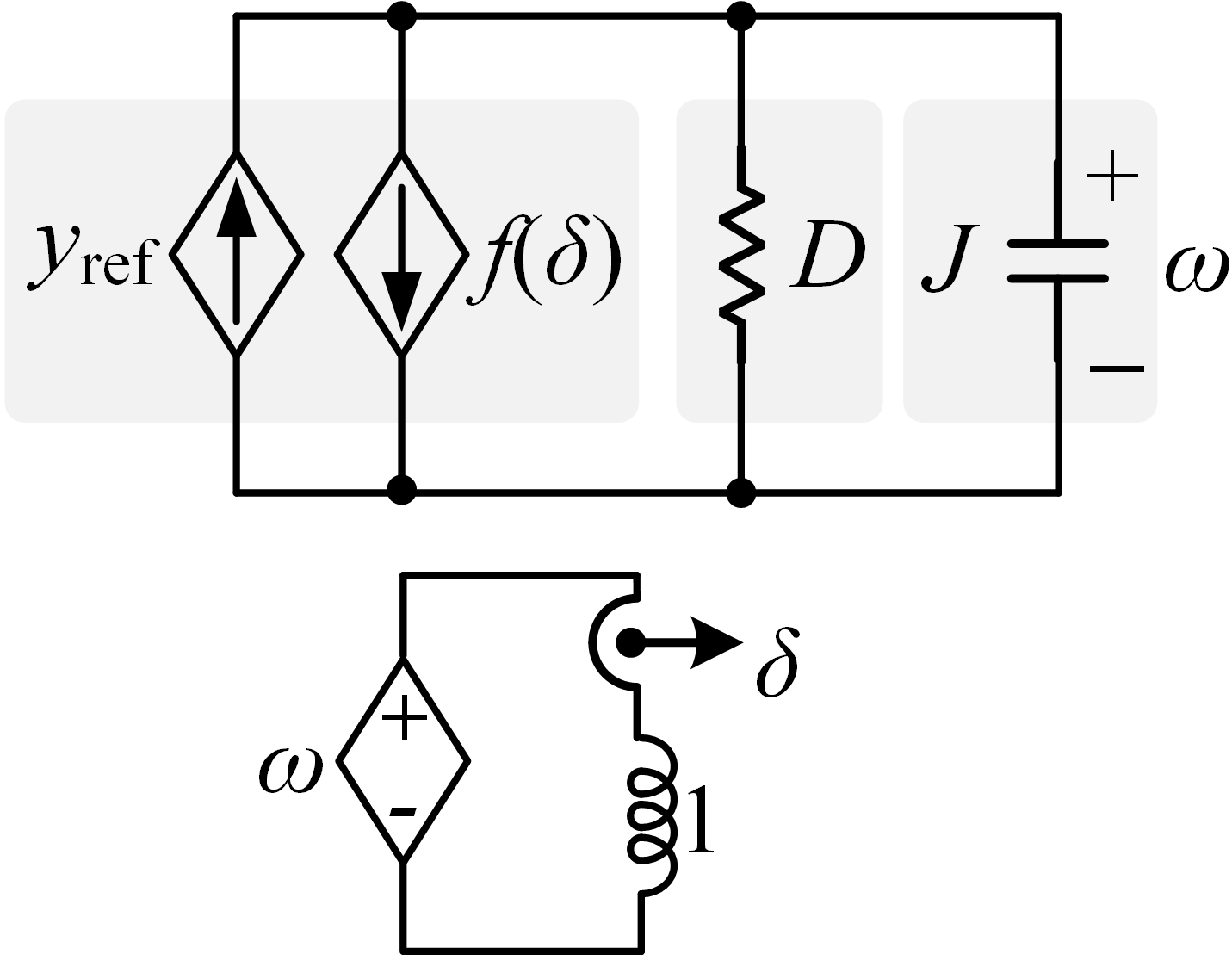}
\end{minipage}
& \begin{minipage}[t]{0.17\textwidth}
\centering
\vspace{-6pt}
\includegraphics[width=\linewidth]{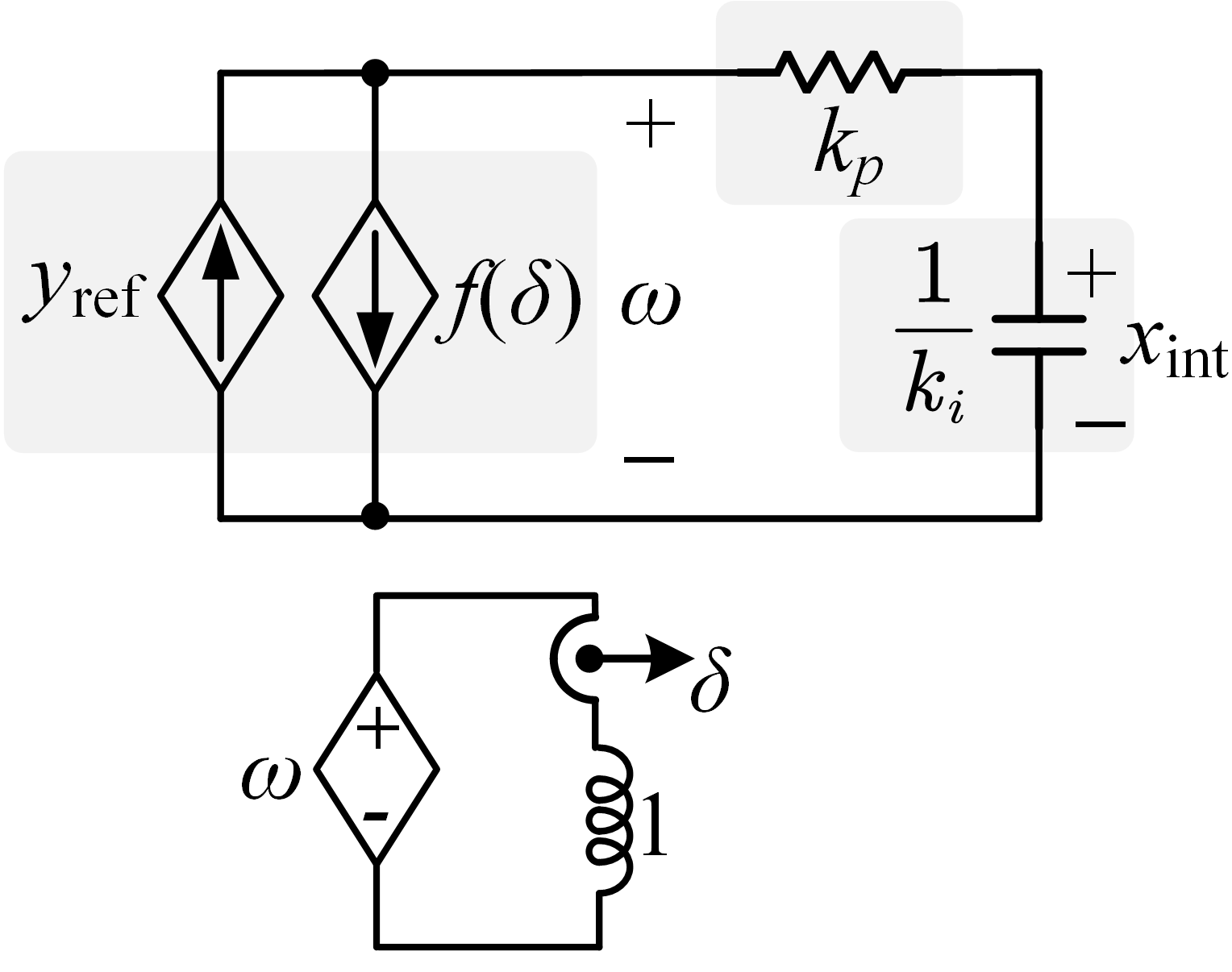}
\end{minipage} \\
\hline\hline
\end{tabular}
}
\vspace{-12pt}
\end{table}
Recalling the GFM inverter or SG, $y_{\mathrm{ref}} = P_{\mathrm{ref}}$ denotes the power reference, and the nonlinear system corresponds to the output power ${EU_g}/{X_g}\sin\delta$. In this case, \eqref{EF_LPF} reduces to the typical form
\[
W=\tfrac{1}{2}J\omega ^2-P_{\mathrm{ref}}\delta -\tfrac{EU_g}{X_g}\cos \delta.
\]

For another widely used controller — the PI controller in \figref{figcontroller}~(b) — the input $y=y_{\text{ref}}-f(\delta)$ is processed by the PI structure rather than the LPF. Let the integrator output be $x_{\text{int}}$. The sum of the proportional and integral terms is then passed through another integrator to drive the nonlinear system $f(\delta)$. The dynamics of \figref{figcontroller}~(b) can be rearranged into a swing-equation form as in prior works:
\vspace{-3pt}
\begin{equation}
    \begin{cases}
	\dot{\delta}=\omega\\
	\underset{J_{\text{eq}}}{\underbrace{\frac{1}{k_i}}}\dot{\omega}=f\left( \delta \right) -\underset{D_{\text{eq}}}{\underbrace{\frac{k_p}{k_i}\frac{\partial f\left( \delta \right)}{\partial \delta}}}\omega\\
\end{cases}
\end{equation}
However, the equivalent damping $D_{eq}$ is not constant but depends on $\tfrac{\partial f(\delta)}{\partial \delta}$. As a result, if the classical energy function is directly applied, the derivative of the energy function becomes indefinite and not necessarily non-positive, owing to the variation of $D_{eq}$.

To overcome this issue, this letter proposes the following energy function for PI-controlled systems:
\begin{equation}
W=\frac{1}{2k_i}{x_{\text{int}}}^2-\int{\left[ y_{\mathrm{ref}}-f\left( \delta \right) \right] d\delta}
\label{EF_PI}
\end{equation}
Compared with \eqref{EF_LPF}, the kinetic energy term $\tfrac{1}{2k_i}x_{\text{int}}^2$ captures the contribution of the PI controller. The derivative of the energy function \eqref{EF_PI} with respect to time is
\begin{equation}
\begin{aligned}
    \dot{W}=-k_p\left[ y_{\mathrm{ref}}-f\left( \delta \right) \right] ^2 \le 0
\end{aligned}
\end{equation}
which is non-positive, and $\{\dot{W}=0\}$ contains no complete trajectory other than the equilibrium. This shows that the PI controller introduces an inherent dissipative mechanism.

The duality between the proposed PI controller energy function and the classical one is further summarized in \tabref{tab:duality}. 
For physical interpretation, equivalent circuits for both energy functions are also illustrated, inspired by \cite{baeckeland2025large}. In both cases, $y_{\text{ref}}$ is modeled as an independent current source and $f(\delta)$ as a controlled current source, whose output energy corresponds to the potential energy term. The quadratic term is associated with the capacitor energy, with $J$ in the classical case and $1/k_i$ in the proposed one, while the dissipation is represented by resistive elements with coefficients $D$ and $k_p$, respectively. Notably, the proposed energy function has a unified form and applies to nonlinear systems of the form shown in \figref{figcontroller}~(b).

\vspace{-6pt}
\section{Representative Applications}
\vspace{-3pt}
This section presents two representative PI-controlled IBR systems to demonstrate ROA estimation and transient stability analysis using the proposed energy function, although the approach is not limited to these cases.
\vspace{-8pt}
\subsection{Case I: PLL of a Grid-Following Inverter}
\begin{figure}[t!]
    \centering
    \vspace{-12pt}
    \includegraphics[width=0.32\textwidth]{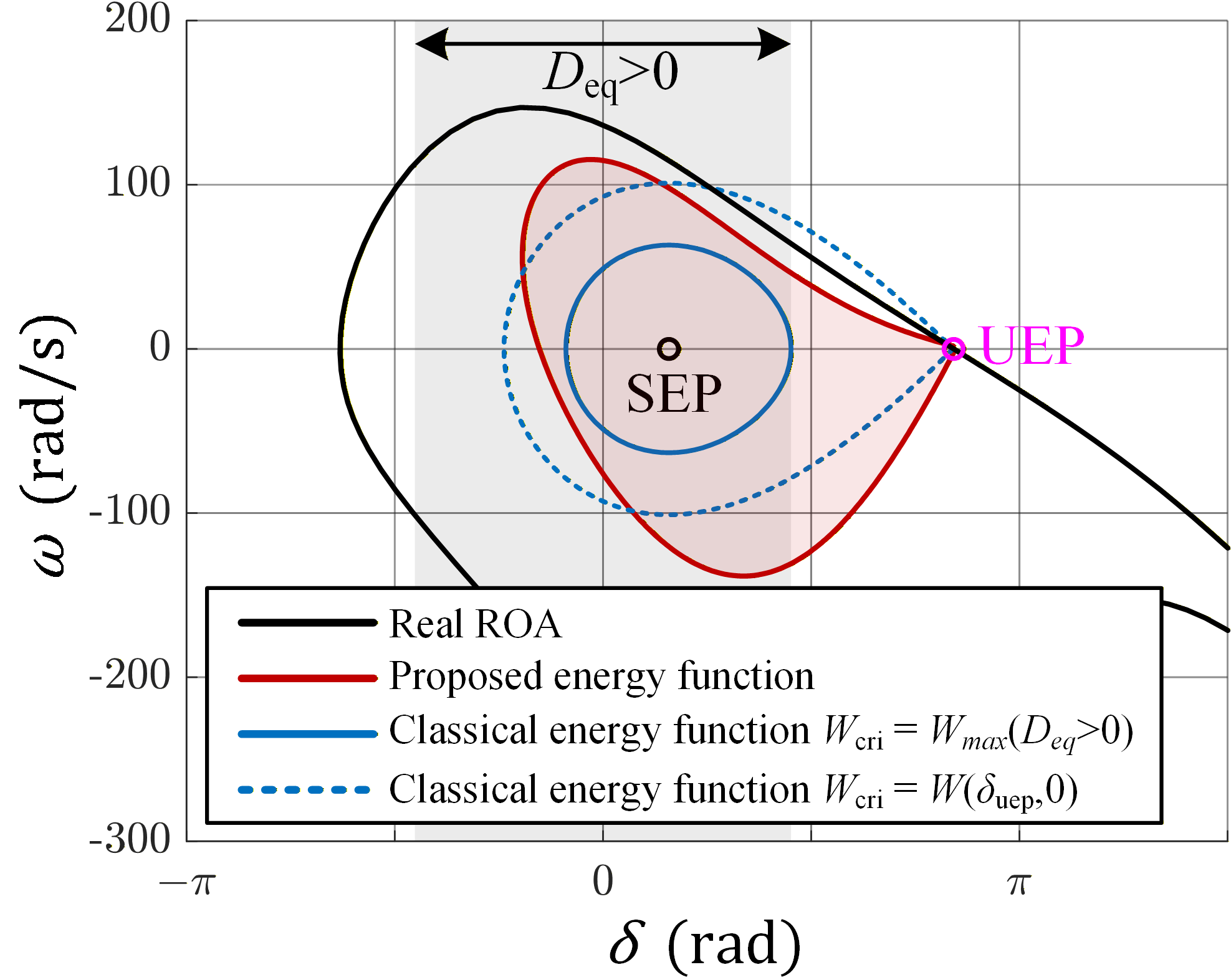}
    \vspace{-9pt}
    \caption{ROA estimation for transient stability of a GFL inverter using the proposed PI-controller energy function and comparison with existing methods.}
    \label{ROApll}	
    \vspace{-9pt}
\end{figure}

A typical example is the PLL in a GFL inverter, whose system and control structures are shown in \figref{ST}~(a)(b) and follow the general form in \figref{figcontroller}~(b). In this case, the PI-controller input $y_{\text{ref}}-f(\delta)$ is the $q$-axis voltage $v_q$, given by
\vspace{-3pt}
\begin{equation}
v_q = X_g I_d + R_g I_q - U_g \sin\delta + \frac{\omega L_g I_d}{\omega_s}
\label{pll1}
\end{equation}
The integrator state is
\vspace{-3pt}
\begin{equation}
x_{\text{int}} = \omega - k_{\text{p,pll}} v_q
\label{pll2}
\end{equation}
Substituting \eqref{pll1} \eqref{pll2} into \eqref{EF_PI} yields the energy function
\vspace{-3pt}
\begin{equation}
\begin{aligned}
    W(\delta,\omega)
    = &\frac{1}{2k_i}\bigl( \omega - k_{\mathrm{p},\mathrm{pll}} v_q \bigr)^2 
    - (X_g I_d + R_g I_q)\delta \\ &- U_g \cos\delta 
    - \int \frac{\omega L_g I_d}{\omega_0} \, d\delta.
    \label{energy_pll}
\end{aligned}
\end{equation}
The ROA is then estimated by the level set $\{(\delta,\omega)\mid W(\delta,\omega)<W_{\mathrm{cri}}\}$, where the critical energy is defined as $W_{\mathrm{cri}}=W(\delta_{\mathrm{uep}},0)$. In \eqref{energy_pll}, the term $\int \tfrac{\omega L_g I_d}{\omega_s}\, d\delta$ is path-dependent introduced by transmission-line dynamics and is commonly approximated along the critical trajectory passing through the UEP \cite{Qu2024}:
\begin{equation}
\vspace{-3pt}
\int_{\mathcal{C}:(\delta_{\mathrm{uep}},0)\rightarrow(\delta,\omega)}
\frac{L_g I_d}{\omega_s}\,\omega\, d\delta
\approx -\frac{\omega(\delta_{\mathrm{uep}}-\delta)}{2}\cdot\frac{L_g I_d}{\omega_s}.
\end{equation}
The estimated ROA is plotted in \figref{ROApll} as the shaded region bounded by the red curve. In comparison, ROA estimation based on classical energy functions, such as \cite{hu2019large} (blue dashed curve) and \cite{fu2020large, mansour2021nonlinear} (blue solid curve), can be inaccurate, because the equivalent damping $D_{\text{eq}}$ may become negative. The proposed method yields a larger stability region without overestimation.

% \begin{figure}[b!]
%     \centering
%     \includegraphics[width=0.3\textwidth]{figures/platform.png}
%     \vspace{-3pt}
%     \caption{HIL experimental platform.}
%     \label{platform}	
%     \vspace{-9pt}
% \end{figure}
\vspace{-8pt}
\subsection{Case II: DC-Voltage-Controlled Grid-Forming Inverter}
\begin{figure}[t!]
\vspace{-12pt}
    \centering
    \includegraphics[width=0.31\textwidth]{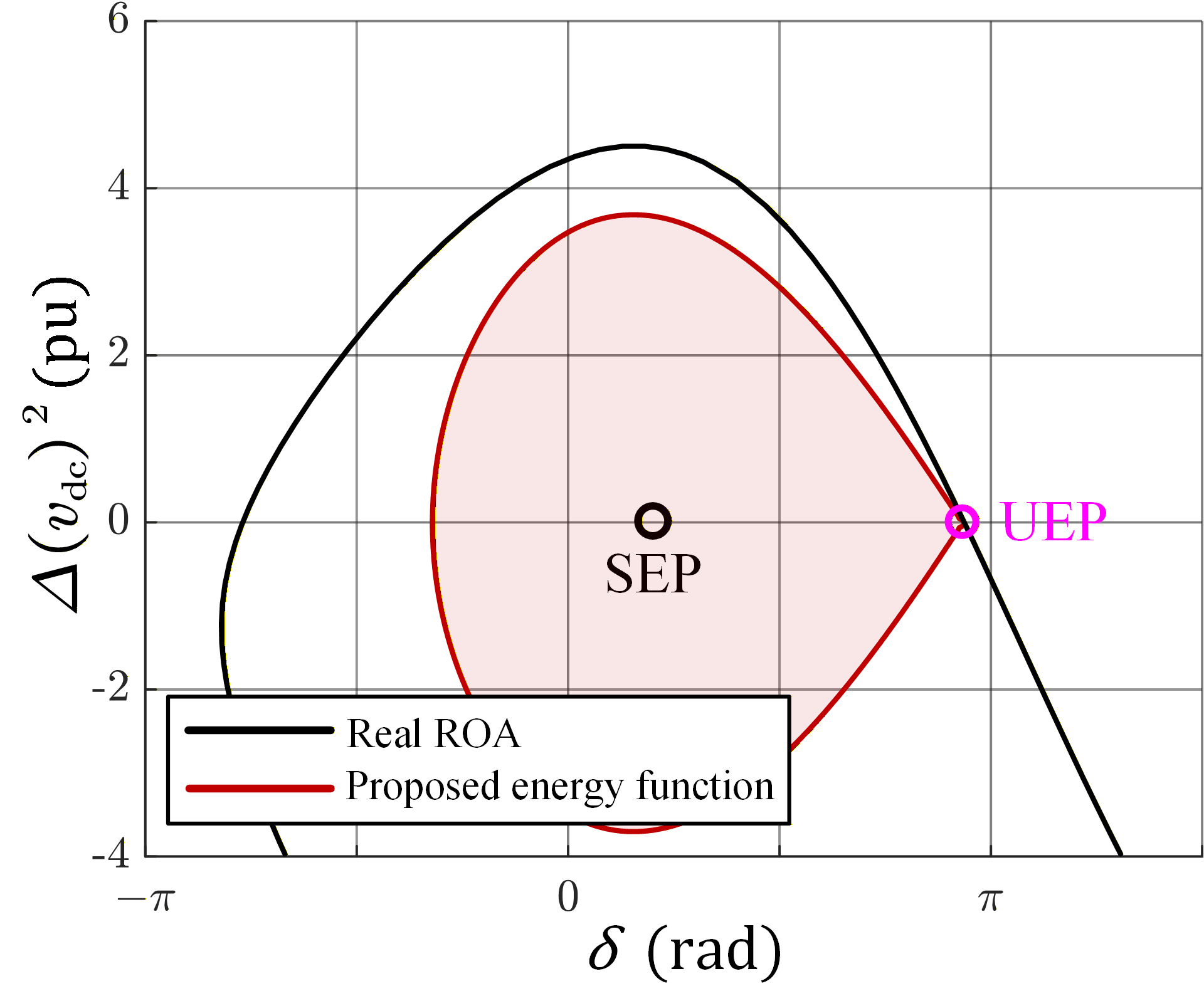}
    \vspace{-9pt}
    \caption{ROA estimation for transient stability of a DC-voltage-controlled GFM inverter using the proposed PI-controller energy function.}
    \label{ROAvfm}	
    \vspace{-9pt}
\end{figure}

The system and control structure are illustrated in \figref{ST}~(a)(c), which also follows the general form in \figref{figcontroller}~(b). Here, the input to the PI controller $y_{\text{ref}}-f(\delta)$ corresponds to the active-power difference between the DC-side input power $P_{\text{in}}$ and the grid-side output power $P_e$
\begin{equation}
\vspace{-3pt}
\begin{aligned}
        P_e=\frac{X_g}{R_{g}^{2}+X_{g}^{2}}EU_g\sin \delta +\frac{R_g}{R_{g}^{2}+X_{g}^{2}}\left( E^2-EU_g\cos \delta \right)
\end{aligned}
\label{vfm1}
\end{equation}
The integrator state is
\begin{equation}
\vspace{-3pt}
x_{\mathrm{int}}=k_i\varDelta (v_{\mathrm{dc}})^2
\label{vfm2}
\end{equation}
where $(\Delta v_{\text{dc}})^2 = v_{\text{dc}}^2 - V_{\text{dc,ref}}^2$ denotes the squared DC-voltage error. Substituting \eqref{vfm1} and \eqref{vfm2} into \eqref{EF_PI} yields
\vspace{-3pt}
\begin{equation}
\footnotesize
\begin{aligned}
    W\left(\delta,\varDelta (v_{\text{dc}})^2 \right) \;=&\; \frac{C}{4}\,k_{i}\big[ \varDelta ( v_{\mathrm{dc}} ) \big]^2 - P_{\text{in}}\delta + \frac{R_g E^2}{R_g^2+X_g^2}\,\delta
       \\
      &- \frac{X_gE U_g}{R_g^2+X_g^2}\, \cos \delta
      - \frac{R_g U_g E}{R_g^2+X_g^2}\,\sin \delta
\end{aligned}
\end{equation}
Similarly, the critical energy is defined as $W_{\mathrm{cri}}=W(\delta_{\mathrm{uep}},0)$, and the ROA is estimated by the level set $\{(\delta,(\Delta v_{\text{dc}})^2)\mid W(\delta,(\Delta v_{\text{dc}})^2)<W_{\mathrm{cri}}\}$. The result is shown in \figref{ROAvfm} (red shaded region). The detailed parameters for all cases are listed in \tabref{TableA}.

\begin{figure}[b!]
    \vspace{-9pt}
    \centering
    \includegraphics[width=0.33\textwidth]{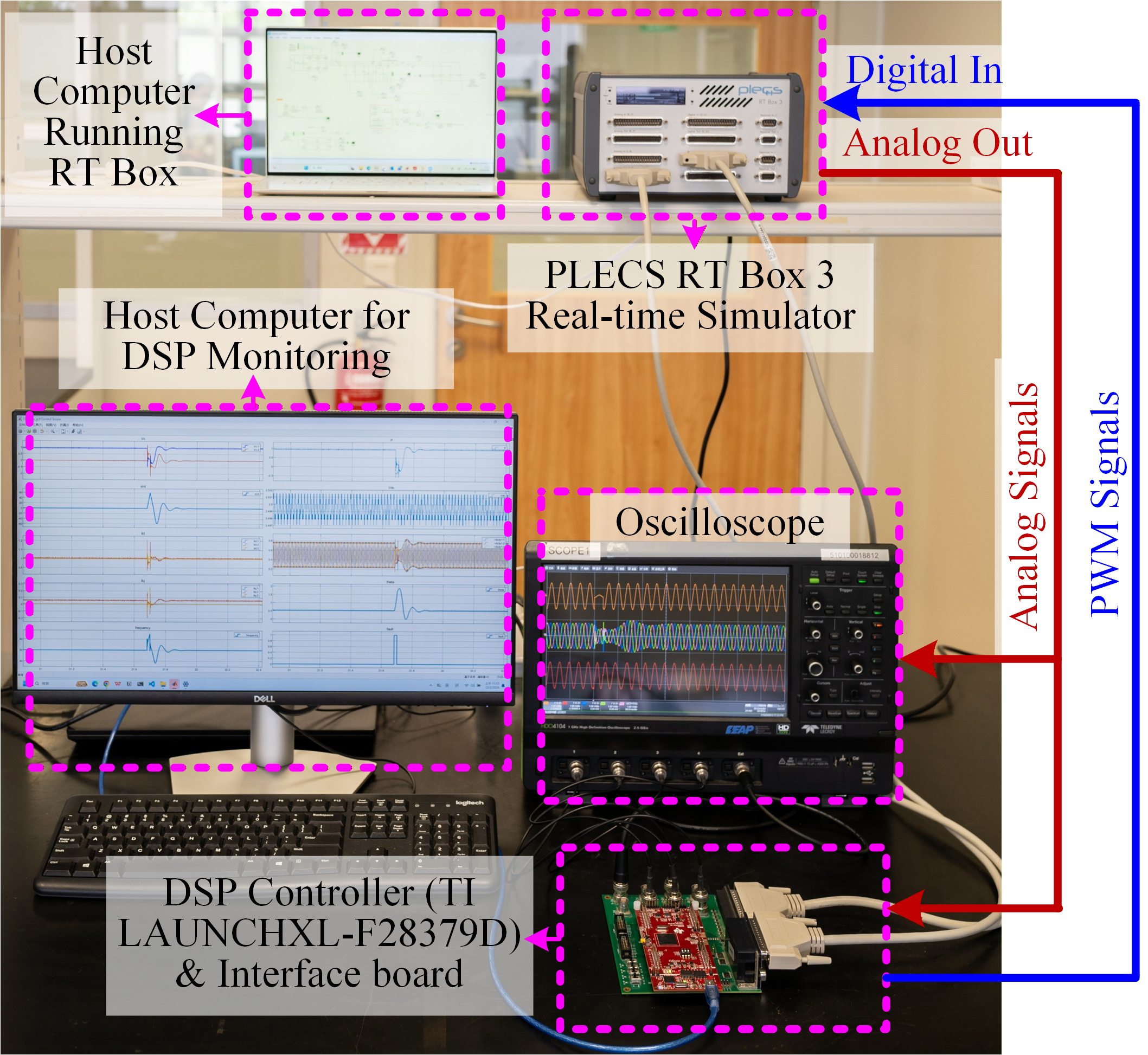}
    \vspace{-6pt}
    \caption{HIL experimental platform.}
    \label{platform}	
    \vspace{-9pt}
\end{figure}

\vspace{-3pt}
\section{HIL Experimental Validation}
A hardware-in-the-loop (HIL) platform is built to validate the proposed PI-controller energy function and the corresponding ROA estimation, as shown in \figref{platform}. The single-inverter infinite-bus system is implemented in a PLECS RT Box 3 real-time simulator, while the controller is executed on a DSP (TMS320F28379D) to generate PWM signals. Detailed parameters are listed in \tabref{TableA}. A voltage sag of 0.1 pu is applied at $t=0$~s.

\vspace{-6pt}
\subsection{Case I: PLL of a Grid-Following Inverter}
\begin{figure*}[t!]
    \centering

    %==================== Left panel: Fig. 6 ====================
    \begin{minipage}[t]{0.33\textwidth}
        \centering
        \captionsetup[subfloat]{labelformat=empty}

        \subfloat[]{\includegraphics[width=\linewidth]{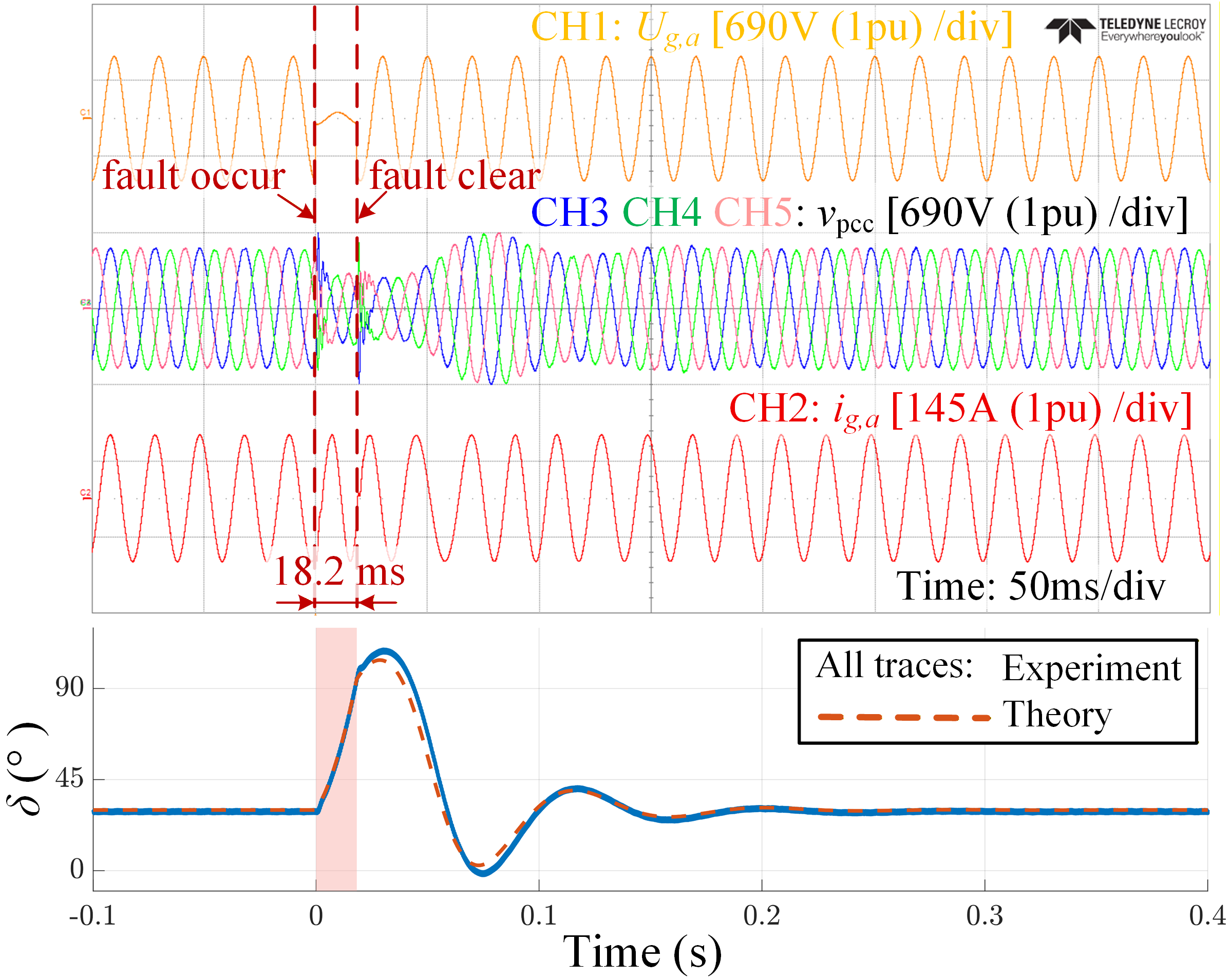}}\\[-8pt]
        {\fontsize{7pt}{8pt}\selectfont
        \parbox{0.98\linewidth}{\raggedright (a)\enspace Time domain when fault duration $t_\text{fault}=18.2$~ms.}}\\[-1pt]

        \subfloat[]{\includegraphics[width=\linewidth]{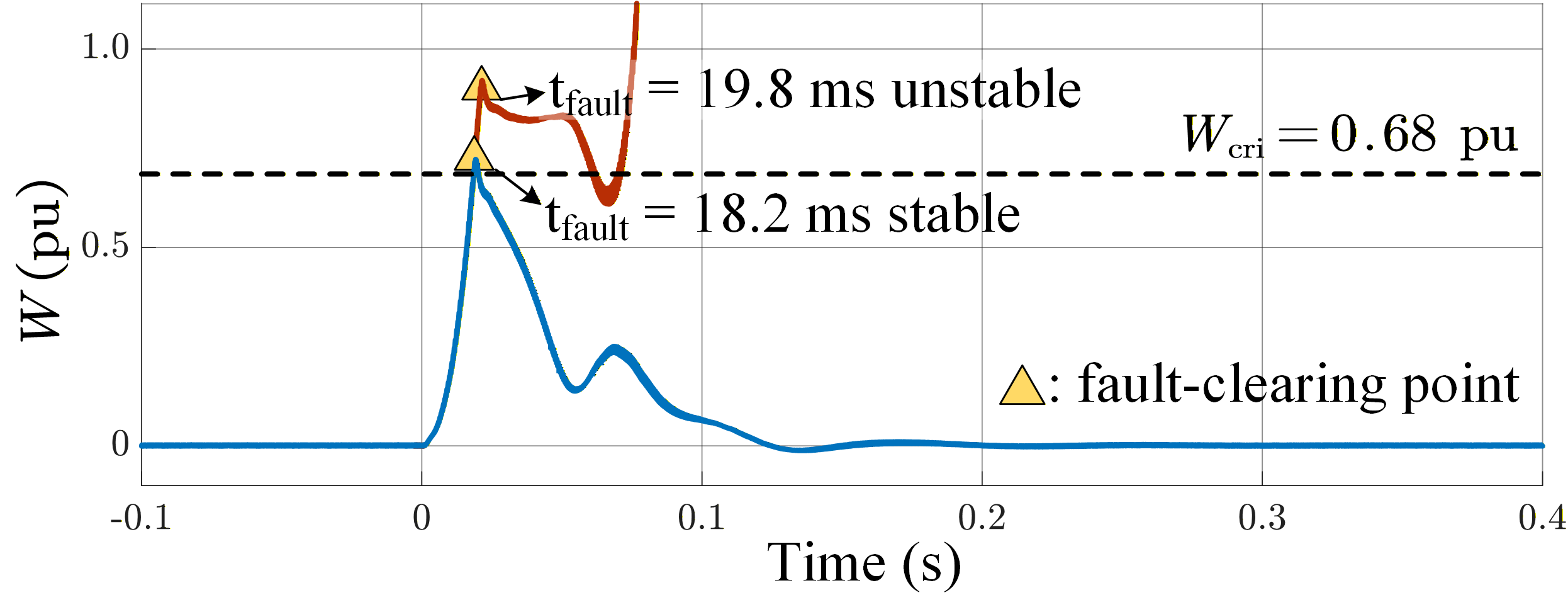}}\\[-8pt]
        {\fontsize{7pt}{8pt}\selectfont
        \parbox{0.98\linewidth}{\raggedright (b)\enspace Value of PI-controller energy function $W$ when fault duration $t_\text{fault}=18.2$~ms and $19.8$~ms respectively (CCT = 19.7~ms).}}
        \vspace{-1pt}
        \caption{Experiment results of GFL inverter.}
        \label{exppll}
    \end{minipage}
    %==================== Middle panel: Fig. 7 ====================
    \begin{minipage}[t]{0.33\textwidth}
        \centering
        \captionsetup[subfloat]{labelformat=empty}

        \subfloat[]{\includegraphics[width=\linewidth]{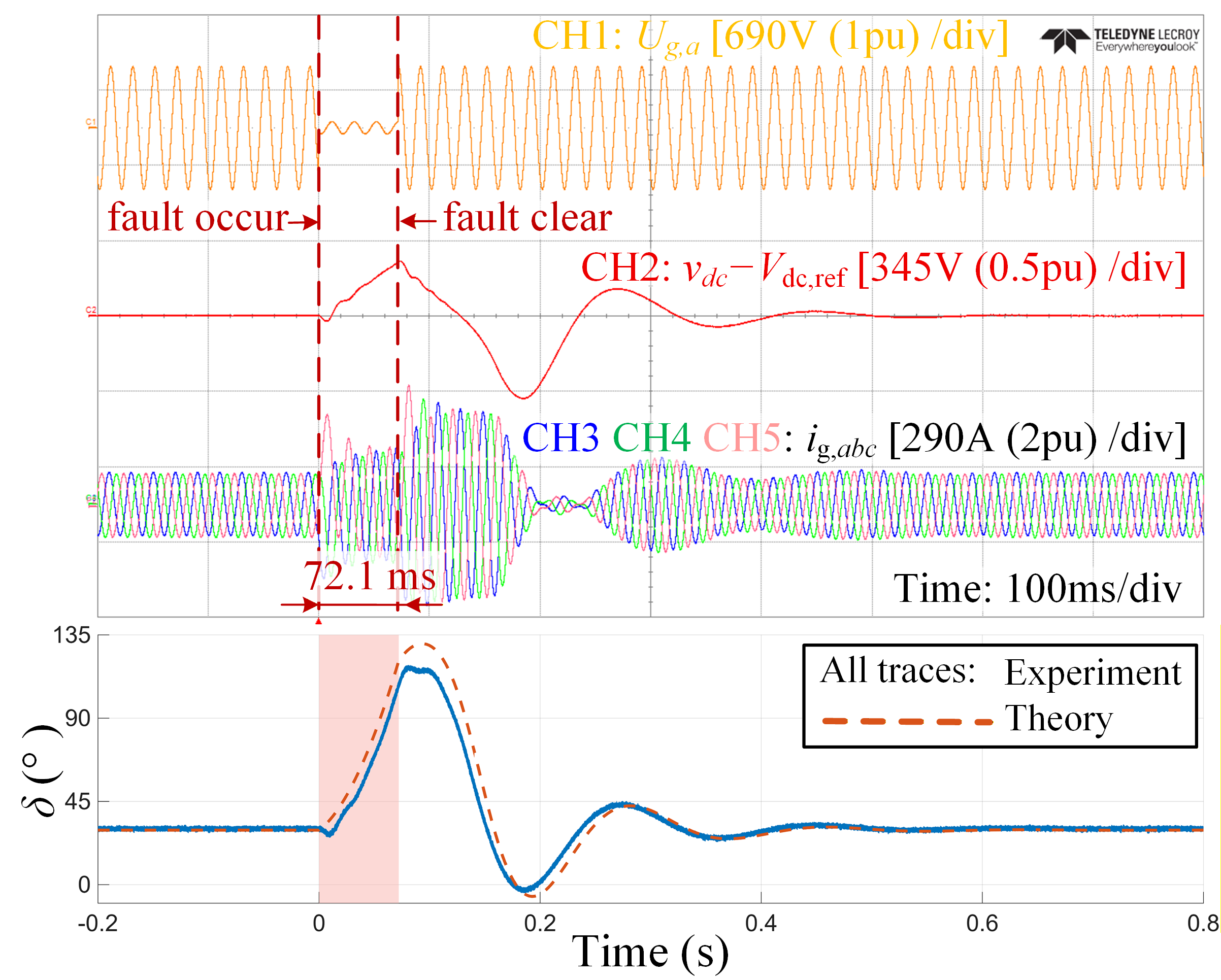}}\\[-8pt]
        {\fontsize{7pt}{8pt}\selectfont
        \parbox{0.98\linewidth}{\raggedright (a)\enspace Time domain when fault duration $t_\text{fault}=72.1$~ms.}}\\[-1pt]

        \subfloat[]{\includegraphics[width=\linewidth]{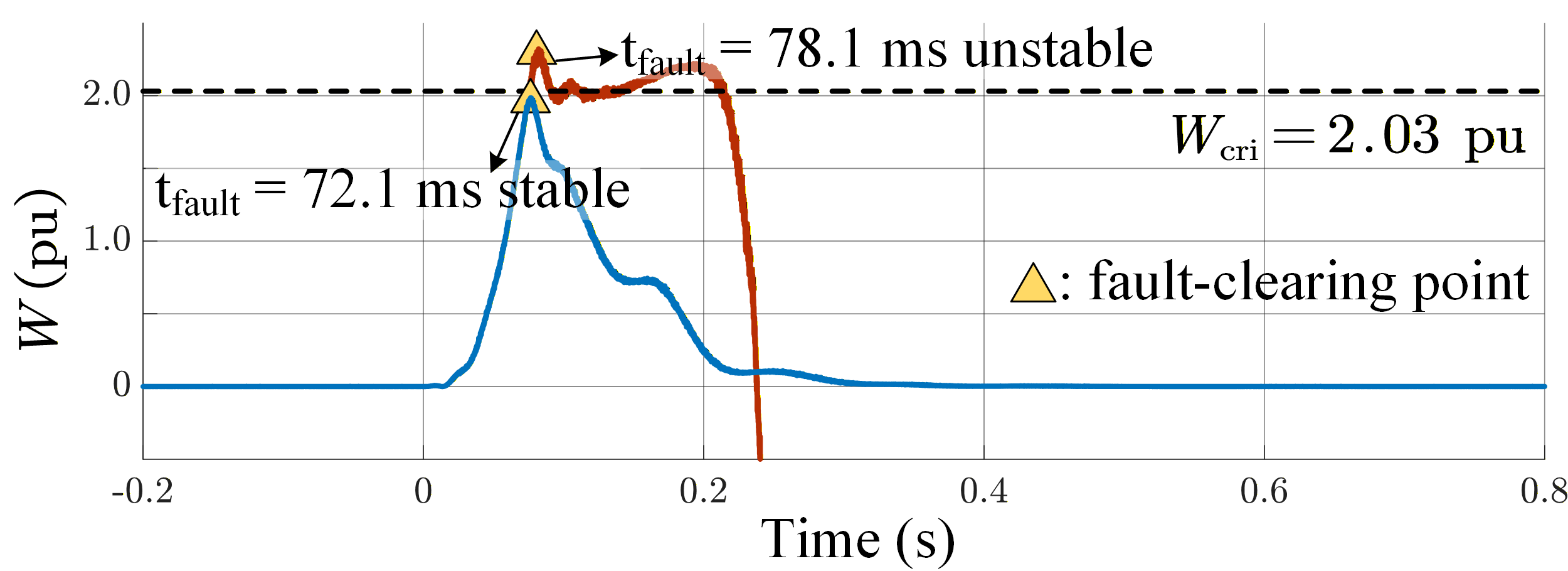}}\\[-8pt]
        {\fontsize{7pt}{8pt}\selectfont
        \parbox{0.98\linewidth}{\raggedright (b)\enspace Value of PI-controller energy function $W$ when fault duration $t_\text{fault}=72.1$~ms and $78.1$~ms respectively (CCT = 78~ms).}}

        \vspace{-1pt}

        \caption{Experiment results of DC-voltage-controller GFM inverter.}
        \label{expvfm}
    \end{minipage}
    \hfill\hfill
    %==================== Right panel: Table A ====================
    \begin{minipage}[t]{0.31\textwidth}
    \centering
    \renewcommand{\thetable}{A1}
    \begin{threeparttable}
    \captionof{table}{Detailed System Parameters}
    \vspace{-24pt}
    \label{TableA}
    \renewcommand{\arraystretch}{1.05}
    \setlength{\tabcolsep}{2pt}
    {\scriptsize
    \everymath{}
    \everydisplay{}
    \begin{tabular}{p{0.58\linewidth} p{0.28\linewidth}}
    \hline\hline
    \multicolumn{1}{l|}{Parameters} & Value\tnote{a} \\ \hline
    \multicolumn{1}{l|}{Grid impedance $Z_g$} & $0.5\mathrm{j}+0.08$~pu \\
    \multicolumn{1}{l|}{Voltage sag depth} & 0.1~pu \\
    \multicolumn{1}{l|}{Grid voltage $U_g$} & 1~pu \\
    \multicolumn{1}{l|}{Base frequency $\omega_0/2\pi$} & 50~Hz \\ \hline
    HIL Setup & \\ \hline
    \multicolumn{1}{l|}{Base capacity $S_b$} & 100~kV$\cdot$A \\
    \multicolumn{1}{l|}{Base voltage $V_b$} & 690~V \\
    \multicolumn{1}{l|}{Base current $I_b$} & 145~A \\
    \multicolumn{1}{l|}{LC filter of Inverter $L_f$} & 0.05~pu \\
    \multicolumn{1}{l|}{LC filter of Inverter $C_f$} & 0.02~pu \\
    \multicolumn{1}{l|}{DC voltage reference $V_{\text{dc,ref}}$} & 2.5~pu \\
    \multicolumn{1}{l|}{Switching \& DSP control frequency} & 10~kHz \\ \hline
    \multicolumn{2}{l}{Case I - PLL in GFL} \\ \hline
    \multicolumn{1}{l|}{PLL controller $k_{\text{p,pll}}$} & $10\times2\pi$~pu \\
    \multicolumn{1}{l|}{PLL controller $k_{\text{i,pll}}$} & $1000\times2\pi$~pu \\
    \multicolumn{1}{l|}{Frequency limit of PLL $\omega_{\text{limit}}$} & $\pm0.3$~pu \\
    \multicolumn{1}{l|}{Current reference $I_d+jI_q$} & 1~pu \\
    \multicolumn{1}{l|}{Inner current control loop bandwidth} & 1~kHz \\ \hline
    \multicolumn{2}{l}{Case II - DC-voltage-controlled GFM} \\ \hline
    \multicolumn{1}{l|}{DC capacitor $C_{\text{dc}}$} & 12.5~pu \\
    \multicolumn{1}{l|}{PI controller $k_p$} & $2\times2\pi$~pu \\
    \multicolumn{1}{l|}{PI controller $k_i$} & 15~pu \\
    \multicolumn{1}{l|}{AC voltage reference $E$} & 1~pu \\
    \multicolumn{1}{l|}{Input power $P_{\text{in}}$} & 1~pu \\
    \multicolumn{1}{l|}{Inner voltage control bandwidth} & 200~Hz \\
    \multicolumn{1}{l|}{Inner current control bandwidth} & 1~kHz \\
    \hline\hline
    \end{tabular}
    }
    \begin{tablenotes}
    \scriptsize
    \item[a] The same per-unit (pu) system is adopted throughout this letter.
    \end{tablenotes}
    \end{threeparttable}
\end{minipage}

    \vspace{-8pt}
\end{figure*}

The critical energy is obtained as $W_{\text{cri}}=0.68$~pu, corresponding to a theoretical CCT of 18.2~ms. The time-domain responses under this fault duration are shown in \figref{exppll}~(a), where the system remains stable and agrees well with the theoretical prediction. The energy trajectory in \figref{exppll}~(b) (blue line) approaches $W_{\text{cri}}$ (black dashed line) at the fault-clearing instant, slightly exceeding it due to deviations between the theoretical model and the practical system. Since the proposed energy function provides a conservative estimate, the result remains conservative. In the HIL experiment, the measured CCT is 19.7~ms. When the fault duration increases to $t_{\text{fault}}=19.8$~ms, the system becomes unstable, as shown by the red energy trajectory in \figref{exppll}~(b), where the energy exceeds $W_{\text{cri}}$ and fails to return to the origin.
% \begin{figure}[t!]
%     \centering

%     \captionsetup[subfloat]{labelformat=empty}
    
%     \subfloat[]{\includegraphics[width=0.38\textwidth]{figures/exp_pll1.png}}\\[-8pt]
%     {\footnotesize\parbox{0.95\linewidth}{\raggedright (a)\enspace Time domain when fault duration $t_\text{fault}=18.2$~ms.}}\\
    
%     \subfloat[]{\includegraphics[width=0.38\textwidth]{figures/exp_pll2.png}}\\[-8pt]
%     {\footnotesize\parbox{0.95\linewidth}{\raggedright (b)\enspace Value of PI-controller energy function $W$ when fault duration $t_\text{fault}=18.2$~ms and $19.8$~ms respectively (CCT = 19.7~ms).}}

%     \caption{Experiment results of GFL inverter.}
%     \label{exppll}	
%     \vspace{-8pt}
% \end{figure}
% \begin{figure}[t!]
%     \centering

%     \captionsetup[subfloat]{labelformat=empty}
    
%     \subfloat[]{\includegraphics[width=0.38\textwidth]{figures/exp_vfm1.png}}\\[-8pt]
%     {\footnotesize\parbox{0.95\linewidth}{\raggedright (a)\enspace Time domain when fault duration $t_\text{fault}=72.1$~ms.}}\\
    
%     \subfloat[]{\includegraphics[width=0.38\textwidth]{figures/exp_vfm2.png}}\\[-8pt]
%     {\footnotesize\parbox{0.95\linewidth}{\raggedright (b)\enspace Value of PI-controller energy function $W$ when fault duration $t_\text{fault}=72.1$~ms and $78.1$~ms respectively (CCT = 78~ms).}}

%     \caption{Experiment results of DC-voltage-controller GFM inverter.}
%     \label{expvfm}
%     \vspace{-8pt}
% \end{figure}
% \vspace{-6pt}
\vspace{-3pt}
\subsection{Case II: DC-Voltage-Controlled Grid-Forming Inverter}
Similarly, the critical energy is obtained as $W_{\text{cri}} = 2.036$~pu, which gives a theoretical CCT of 72.1~ms. Under this fault duration, the time-domain responses are shown in \figref{expvfm}~(a). It is shown that the system remains stable, and the HIL experimental results closely match the theoretical model. The value of the PI energy function is shown in \figref{expvfm}~(b)  (blue solid line). At the fault-clearing instant, the energy is nearly equal to $W_{\text{cri}}$ (black dashed line), but slightly higher due to experimental imperfections.  Since the proposed energy function provides a conservative estimate, the result remains conservative. In the HIL experiment, the measured CCT is 78~ms. An additional experiment is performed with $t_{\text{fault}}=78.1$~ms, where the system becomes critically unstable. The corresponding energy trajectory is also shown in \figref{expvfm}~(b) (red solid line). When the fault duration is too large and  the energy rises above $W_{\text{cri}}$, the system loses stability and the energy cannot return to the origin.

\vspace{-6pt}
\section{Conclusions}
This letter proposes a unified PI-controller energy function for transient synchronization stability analysis of IBRs. The proposed formulation reveals the inherent dissipation mechanism introduced by the PI controller, leading to a non-positive energy derivative.  The method is not limited to the GFL inverter or the DC-voltage-controlled GFM inverter, which is investigated in the letter, but is generally applicable to a class of IBR systems employing PI-based synchronization structures. The effectiveness of the proposed approach and the validity of the analysis are further confirmed by HIL experiments.

\vspace{-0.1cm}

%% file: appendix.tex
%\appendices
\section*{Appendix}\label{A}
\vspace{-3pt}
\setcounter{table}{0}
\setcounter{figure}{0}
\renewcommand{\thetable}{A\arabic{table}}
\renewcommand\cellalign{tl}
\renewcommand{\theequation}{A\arabic{equation}}
\renewcommand{\thefigure}{A\arabic{figure}} 
The detailed parameters are listed in \tabref{TableA}.